\documentclass[draft,nofootinbib,pre,twocolumn,showpacs,showkeys,preprintnumbers,floatfix]{revtex4-1}

\usepackage{dynlearn}
\usepackage{hyperref}
\usepackage{graphicx}
\usepackage{rotating}

\usepackage{physics}
\usepackage{multibib}
\usepackage{xcolor}
\newcites{SM}{SM References}
\usepackage{mathrsfs}
\usepackage[english]{babel}
\usepackage{amssymb,amscd}

\newcommand{\CC}[2]{\mathcal{C}_{#1 #2}}

\newcommand{\MV}[1]{\left\langle #1 \right\rangle}

\newcommand{\intP}{\int_{\mathcal{P}(\mathcal{H})} \!\!\!\!\!\!\!\!\!}

\newcommand{\PH}{\mathcal{P}(\mathcal{H})}

\newcommand{\p}{\partial}

\def\tbf #1 {\textbf{#1} }

\begin{document}

\def\ourTitle{%
A kinetic theory for quantum information transport
}

\def\ourAbstract{%
In this work we build a theoretical framework for the transport of information in quantum systems. 
This is a framework aimed at describing how out of equilibrium open quantum systems move information 
around their state space, using an approach inspired by transport theories. The main goal is to build new 
mathematical tools, together with physical intuition, to improve our understanding of non-equilibrium 
phenomena in quantum systems. In particular, we are aiming at unraveling the interplay between 
dynamical properties and information-theoretic features. The main rationale here is to have a framework 
that can imitate, and potentially replicate, the decades-long history of success of transport theories in 
modeling non-equilibrium phenomena. 
}

\def\ourKeywords{%
Quantum Mechanics, Geometric Quantum Mechanics
}

\hypersetup{
  pdfauthor={Fabio Anza},
  pdftitle={\ourTitle},
  pdfsubject={\ourAbstract},
  pdfkeywords={\ourKeywords},
  pdfproducer={},
  pdfcreator={}
}


\title{\ourTitle}

\author{Fabio Anza}
\email{fanza@ucdavis.edu}


\affiliation{Complexity Sciences Center and Physics Department,
University of California at Davis, One Shields Avenue, Davis, CA 95616}

\date{\today}
\bibliographystyle{unsrt}

\begin{abstract}
\ourAbstract
\end{abstract}



\preprint{\arxiv{2008.08682}}

\date{\today}
\maketitle





\section*{Introduction.} The study of transport phenomena is crucial to harness 
the dynamical properties of natural systems. In this sense, transport theories are 
simplified phenomenological descriptions of the nonequilibrium behavior of a system. 
Well-known examples are the theories of transport of charge, mass and heat. Most of them were 
originally formulated to understand macroscopic phenomena like, in the case of 
mass transport, the tendency of a system to transfer mass to minimize the 
concentration difference. Over time, they were put on more rigorous ground using 
statistical mechanics and kinetic theory, and their use has been incredibly powerful in understanding
and modeling emergent non-equilibrium properties of many-particles systems. Most notably, 
Boltzmann used this approach to describe the behavior of gases, and explain the rise of 
macroscopic irreversibility from the underlying time-symmetric classical mechanics. After 
Boltzmann's work, the core idea of treating motion and interactions in a statistical way has 
led to innumerable advances, both of fundamental and applied nature. With the rise of 
quantum theory, these techniques have also been adapted to include quantum fluctuations, 
leading to improved descriptions of transport phenomena at the nanoscale. 
While nowadays the use of kinetic and transport theories is ubiquitous, and also appropriate
at various length scales (from the lower nanoscale to the larger scale of stars and galaxies), 
these techniques have not been leveraged to study quantum information in closed and open 
quantum systems. Thus, in this work, we attempt to lay a bridge between the field of quantum 
information and that of transport theories, with the goal of building new tools to investigate the
interplay between non-equilibrium physical properties of quantum systems and their information-theoretic
features. More accurately, here we provide a self-consistent framework to track and model the 
dynamical evolution of a quantum state (and its probability distribution) in nonequilibrium quantum 
systems. We do so by building a kinetic theory of quantum state evolution, on the quantum 
state space, both in closed and open configurations.

At the technical level, this is done by leveraging recent work \cite{Anza20a,Anza20b} on 
Geometric Quantum Mechanics: a differential-geometric approach to quantum mechanics that 
gets rid of the phase redundancy of the Hilbert space and exposes the underlying phase-space-like
structure of the space of quantum states, with probabilities and phases playing the role of 
canonically conjugated coordinates.

The paper is organized as follows. In Section \ref{sec:GQM} we give a brief summary of 
GQM, and of the tools introduced in \cite{Anza20a,Anza20b}, both of which are needed to
follow the details of the main part of the work. In Section \ref{sec:IT} we give our first result: 
a microscopic derivation of the continuity equation for the information transport in quantum 
systems, with the appropriate microscopic definitions for phenomenological quantities such 
as information flux and sink/sources. Section \ref{sec:DYN} contains our second result: a 
general equation of motion for the probability distribution to find an open quantum system 
in one of its pure states. In Sections \ref{sec:EXAMPLES1} and \ref{sec:EXAMPLES2} 
we use the theoretical framework developed to analyze a few concrete examples, supported 
by numerical analysis. Eventually, in Section \ref{sec:FINAL} we draw some conclusions.



\section{Geometric Quantum Mechanics}
\label{sec:GQM}

References
\cite{STROCCHI1966,Miel68,Kibble1979,Heslot1985,Page87,And90,Gibbons1992,Ashtekar1995,Ashtekar1999,Brody2001,Bengtsson2017,Carinena2007,Chruscinski2006,Marmo2010,Avron2020,Pastorello2015,Pastorello2015a,Pastorello2016,Clemente-Gallardo2013}
give a comprehensive introduction to GQM. Here, we briefly summarize only the
elements we need, working with Hilbert spaces $\mathcal{H}$ of finite dimension $D$.
For the details of the derivations, we send the reader to the literature cited above.

Given an arbitrary basis $\left\{\ket{e_n} \right\}_{n=0}^{D-1}$, a pure state is
parametrized by $D$ complex homogeneous coordinates $Z = \left\{      Z^n\right\}$, up to
normalization and an overall phase:
\begin{align*}
\ket{\psi} = \sum_{n=0}^{D-1} Z^n \ket{e_n}~.
\end{align*}
Here, and throughout the paper, we will always use upper indices to identify different 
coordinates of the same point and lower indices to identify different point. This 
description is redundant since we can renormalize $Z$ by multiplying it for a real number
and a phase (hence a complex number) and we would get the same physical state. Therefore, 
$Z \in \mathbb{C}^{D}$, $Z \sim \lambda Z$, with $\lambda \in \mathbb{C}/\left\{ 0\right\}$. This equivalence 
relation means pure states of a quantum system are points in the complex projective space $\mathcal{P}\left(
\mathcal{H} \right)=\mathbb{C}\mathrm{P}^{D-1}$. We will often refer to $\PH$ as the \emph{quantum 
state space}. One can always use probability-phase coordinates which, in the case of a single qubit
are $Z = (\sqrt{1-p},\sqrt{p} e^{i\phi})$. We will often refer to this set of coordinates as 
they play a particular useful role.

\paragraph*{Observables and POVMs.} In GQM, an \emph{observable} is a function $\mathcal{O}(Z) \in
\mathbb{R}$ that associates to each point  $Z \in \mathcal{P}(\mathcal{H})$ the
expectation value $\bra{\psi} \mathcal{O} \ket{\psi}/\braket{\psi}{\psi}$ of the corresponding
operator $\mathcal{O}$ on state $\ket{\psi}$ with coordinates $Z$:
\begin{align}
\mathcal{O}(Z) = \frac{\sum_{\alpha,\beta} \mathcal{O}_{\alpha,\beta}Z^\alpha \overline{Z}^\beta}{\sum_{\gamma} \left\vert Z^\gamma\right\vert^2}
  ~,
\label{eq:GQM_Observable}
\end{align}
where $\mathcal{O}_{\alpha \beta}$ is Hermitian $\mathcal{O}_{\beta,\alpha} = \overline{\mathcal{O}}_{\alpha,\beta}$.

Measurement outcome probabilities are determined by \emph{positive
operator-valued measurements} (POVMs) $\left\{E_j\right\}_{j=1}^D$ applied to a
state \cite{Nielsen2010,Heinosaari2012}. They are nonnegative operators
$E_j\geq 0$, called \emph{effects}, that sum up to the identity: $\sum_{j=1}^{D}
E_j = \mathbb{I}$. In GQM they consist of nonnegative real functions $E_j(Z)\ge
0$ on $\mathcal{P}(\mathcal{H})$ whose sum is always unity:
\begin{align}
E_j(Z) = \frac{\sum_{m,n}
  \left(E_j\right)_{m,n} Z^m \overline{Z}^n}{\sum_{k} \left\vert Z^k \right\vert^2}
  ~,
\label{eq:GQM_POVMs}
\end{align}
where $\sum_{j=1}^{D}E_j(Z) = 1$.

The quantum state space $\mathcal{P}(\mathcal{H})$ has a preferred metric 
$g_{FS}$---the \emph{Fubini-Study metric} \cite{Bengtsson2017}---and an 
associated volume element $dV_{FS}$ that is coordinate-independent and
invariant under unitary transformations. The geometric derivation of $dV_{FS}$
is beyond our immediate goals here. That said, it is sufficient to give its
explicit form in the probability-phase coordinate system $Z^n =
\sqrt{p_n}e^{i\phi_n}$ that we are going to use for explicit calculations: 
\begin{align*}
dV_{FS}
  & = \sqrt{\det g_{FS}}
  \prod_{n=0}^{D-1} dZ^n d\overline{Z}^n \\
  & =  \prod_{n=1}^{D-1} \frac{dp_n d\phi_n}{2}
  ~.
\end{align*}
Notice how $p_0$ and $\phi_0$ are not involved. This is due to
$\mathcal{P}(\mathcal{H})$'s projective nature which guarantees that we can
choose a coordinate patch in which $p_0 = 1 - \sum_{n=1}^{D-1}p_n$
and $\phi_0 = 0$. As we see now, the probability-phase coordinates play a 
particular role: they are canonically conjugated.

\paragraph*{State-space structure.} Beyond the Riemannian structures, it can 
be shown that $\PH$ also has another interesting geometric feature: a symplectic structure.
This is the hallmark of classical state spaces and justifies the use of the term 
\emph{quantum state space} for $\PH$. With a more common jargon, a symplectic
structure is the geometric entity allowing us to define ``Poisson Brackets''
and the existence of canonically conjugated coordinates. In particular, using 
probabilities and phases coordinates $\left\{ (p_n,\phi_n)\right\}$
one has that $\left\{ p_n , \phi_n \right\} = \frac{1}{\hbar}\delta_{nm}$. Thus,
for arbitrary functions $A$ and $B$ on $\PH$ one has
\begin{equation}
\left\{ A, B\right\} \coloneqq \frac{1}{\hbar}\sum_{n=1}^{D-1} \frac{\partial A}{\partial p_n} \frac{\partial B}{\partial \phi_n} - \frac{\partial B}{\partial p_n} \frac{\partial A}{\partial \phi_n}
\end{equation}

\paragraph*{Unitary evolution.} In QM, an isolated quantum system evolves
with a unitary propagator $U(t,t_0) = e^{-\frac{i}{\hbar}H (t-t_0)}$, where the
generator $H$ is the (time-independent) Hamiltonian of the system. Surprisingly, 
it can be shown \cite{Bengtsson2017} that this evolution is equivalent to a classical 
Hamiltonian dynamics, with geometric coordinates. In particular, calling $E(p_n, \phi_n) = \bra{\psi(p_n,\phi_n)}H \ket{\psi(p_n,\phi_n)}$
the expectation value of $H$ on a generic state $\ket{\psi(p_n,\phi_n)} = \sum_{n=0}^{D-1}\sqrt{p_n}e^{i\phi_n}\ket{e_n}$
parametrized by $(p_n,\phi_n)_n$, the unitary evolution of a generic function $A$
on $\PH$ is given by 
\begin{equation}
\frac{\partial A}{\partial t} = \left\{ A,E\right\}
\end{equation}
Or, equivalently, a generic state $(p_n,\phi_n)_n$
evolves according to Hamilton's equations of motion:
\begin{subequations}\label{eq:HAM_EOM}
\begin{align}
&\frac{dp_n}{dt} = \frac{1}{\hbar}\frac{\partial E}{\partial \phi_n} \\
&\frac{d\phi_n}{dt} = -\frac{1}{\hbar}\frac{\partial E}{\partial p_n} 
\end{align}
\end{subequations}

Here the analogies with classical Hamiltonian mechanics are particularly evident.
While quantum mechanics is clearly very different from its classical
counterpart, at a certain descriptive level we can still use the intuition
of classical mechanics to understand the dynamical evolution of quantum systems. 
We now proceed in our summary by looking at situations in which the state of the 
system can not be characterized by a single point, thus needing a probability distribution 
over the whole state space.

\paragraph*{Geometric quantum states.}
The geometric framework makes it very natural to view a quantum state as a functional
encoding that associates expectation values to observables, paralleling the
$C^{*}$-algebra formulation of quantum mechanics \cite{Strocchi2008a}. 
The idea is that one considers probability density functions on $\mathcal{P}(\mathcal{H})$,
together with observable functions. 

Geometric Quantum States are functionals $P_q[\mathcal{O}]$ from the algebra of
observables $\mathcal{A}$ to the real line: 
\begin{align}
P_q[\mathcal{O}]
  = \int_{\mathcal{P}(\mathcal{H})} q(Z) \mathcal{O}(Z) dV_{FS}
  ~,
\label{eq:gqs}
\end{align}
where $\mathcal{O} \in \mathcal{A}$, $q(Z) \geq 0$ is the
normalized distribution associated with the functional $P_q$:
\begin{align*}
P_q[1] = \int_{\mathcal{P}(\mathcal{H})}
  q(Z) dV_{FS}  = 1
  ~,
\end{align*}
and $P_q[\mathcal{O}] \in \mathbb{R}$. 
Thus, kets $\ket{\psi_0}$ are described by functionals with a Dirac-delta
distribution $q_0(Z) = \widetilde{\delta}\left[ Z - Z_0\right]$:
\begin{align*}
P_{0}[\mathcal{O}] &= \intP \widetilde{\delta}(Z-Z_0)\mathcal{O}(Z) dV_{FS} \\
  & = \mathcal{O}(Z_0)  = \bra{\psi_0}\mathcal{O}\ket{\psi_0}
  ~.
\end{align*}
Here, $\widetilde{\delta}(Z-Z_0)$ is shorthand for a coordinate-covariant Dirac-delta in
arbitrary coordinates. In homogeneous coordinates this reads:
\begin{align*}
\widetilde{\delta}(Z - Z_0) \coloneqq \frac{1}{\sqrt{\det g_{FS}}}
  \prod_{n=0}^{D-1} \delta(X^n - X^n_0)
  \delta(Y^n - Y^n_0)
  ~,
\end{align*}
where $Z^n = X^n + iY^n$. In $(p_n,\phi_n)$ coordinates
this becomes simply:
\begin{align*}
\widetilde{\delta}(Z - Z_0) = \prod_{n=1}^{D-1} 2\delta(p_n - p_n^0) \delta(\phi_n - \phi_n^0)
  ~,
\end{align*}
where the coordinate-invariant nature of the functionals $P_q[\mathcal{O}]$ is
now apparent. Extending by linearity, a specific decomposition $\left\{ \lambda_j, \ket{\lambda_j}\right\}$
of a density matrix
\begin{align*}
\rho = \sum_{j}\lambda_j \ket{\lambda_j}\bra{\lambda_j}
\end{align*}
corresponds to a convex combinations of these Dirac-delta functionals:
\begin{align*}
q_{\mathrm{mix}}(Z) = \sum_{j}\lambda_j \widetilde{\delta}(Z-Z_j)
  ~. 
\end{align*}
Thus, expressed as functionals from observables to the real line, mixed states
are:
\begin{align}
P_{\mathrm{mix}}\left[ \mathcal{O}\right]
  & = \sum_{j} \lambda_j \bra{\lambda_j}\mathcal{O}\ket{\lambda_j}
  ~.
\label{eq:Functional}
\end{align}
 
Equipped with this tools, one identifies the distribution $q(Z)$ of Eq.
(\ref{eq:gqs}) as a system's \emph{geometric quantum state}. This is a
generalized notion of quantum state.

A simple example of a geometric quantum state is the 
\emph{geometric canonical ensemble}:
\begin{align*}
q(Z) = \frac{1}{Q_\beta} e^{-\beta h(Z)}
  ~,
\end{align*}
where:
\begin{align*}
  Q_\beta & = \int dV_{FS} e^{-\beta h(Z)} ~, \\
  h(Z) & = \bra{\psi(Z)}H\ket{\psi(Z)} ~,
\end{align*}
and $H$ is the system's Hamiltonian operator. This was introduced in 
Refs. \cite{Brody1998}. References \cite{Brody2016} and \cite{Anza20b} 
investigated its potential role in establishing a quantum foundation of
thermodynamics that is an alternative to that based on Gibbs ensembles and von
Neumann entropy. 

\paragraph*{Density matrix.}
The connection between geometric quantum states and density matrices is
fairly straightforward. Since density matrices are a synthetic way of collecting
probability outcomes about POVMs, which are functions of the 
form $A(Z) \propto \sum_{n,m}A_{n m} Z^n \overline{Z}^m$, 
given a generic geometric quantum state $q(Z)$ the associated density matrix $\rho^q$ can be 
simply computed as follows:
\begin{align}
\rho^q_{mn} & = P_q[Z^m \overline{Z}^n] \nonumber \\
  & = \intP dV_{FS} \, q(Z)  \, Z^m \overline{Z}^n
  ~.
\label{eq:densitymatrix}
\end{align}
Owing to the specific form of POVMs on $\mathcal{P}(\mathcal{H})$, recall Eq. (\ref{eq:GQM_POVMs}), they
are sensitive to $q(Z)$ only via $\rho^q$. Therefore, if two geometric quantum states 
$q_1$ and $q_2$ induce the same density matrix $\rho^{q_1} = \rho^{q_2}$, then all POVMs
produce the same outcomes.

\paragraph*{The GQS of an Open Quantum System.}
As shown above, if a system is in a pure state its GQS is simply a Dirac delta concentrated
on a single point. However, if the system is in contact with an environment, this is not true 
anymore, unless they are exactly in a product state. In general, contact with the environment
causes the loss of information about which region of the state space the system inhabits. Hence, 
we need a probability distribution for the state of our system: a geometric quantum state. The explicit 
derivation of the generic geometric quantum state for an open quantum system can be found 
in Ref.\cite{Anza20a}. Here we simply give the result, without discussing the derivation. 
Given a system and its environment, we call $\left\{ \ket{a_j}\right\}_{j=0}^{d_S-1}$ a basis for 
the Hilbert space $\mathcal{H}_S$ of the system (dimension $d_S$) and 
$\left\{ \ket{b_n}\right\}_{n=0}^{d_E}$ a basis for the Hilbert space $\mathcal{H}_E$ 
of the environment (dimension $d_E$). Assuming system and environment are in 
a pure state $\ket{\psi_{SE}} = \sum_{j,\alpha}\psi_{j\alpha}\ket{a_j}\ket{b_\alpha}$, the 
geometric quantum state of the system is
\begin{equation}
q(Z) = \sum_{\alpha=0}^{d_E-1} x_\alpha \widetilde{\delta}\left( Z-Z(\chi_\alpha)\right)~,\label{eq:gqs_opn}
\end{equation}
where 
\begin{subequations}\label{eq:def_xchi}
\begin{align}
&x_\alpha \coloneqq \sum_j \left\vert \psi_{j\alpha}\right\vert^2~,\\
&Z^j(\chi_\alpha) = \frac{\psi_{j\alpha}}{\sqrt{x_\alpha}}\quad \to \quad \ket{\chi_\alpha} = \sum_j \frac{\psi_{j\alpha}}{\sqrt{x_\alpha}}\ket{a_j}
\end{align}\label{eq:xgammas}
\end{subequations}
Here $x_\alpha$ is the probability that the environment is in state $\ket{b_\alpha}$ and $\left\{\ket{\chi_\alpha}\right\}$ 
is a set of $d_E$ states of the system which constitutes the discrete support of the geometric 
quantum state. The geometric quantum state in Eq.(\ref{eq:gqs_opn}) provides the correct reduced density matrix of the system
$\rho^{S}(\psi_{SE}) = \sum_\alpha x_\alpha \ket{\chi_\alpha}\bra{\chi_\alpha}$.
\begin{figure*}[t!]
\centering
\includegraphics[width=.9\textwidth]{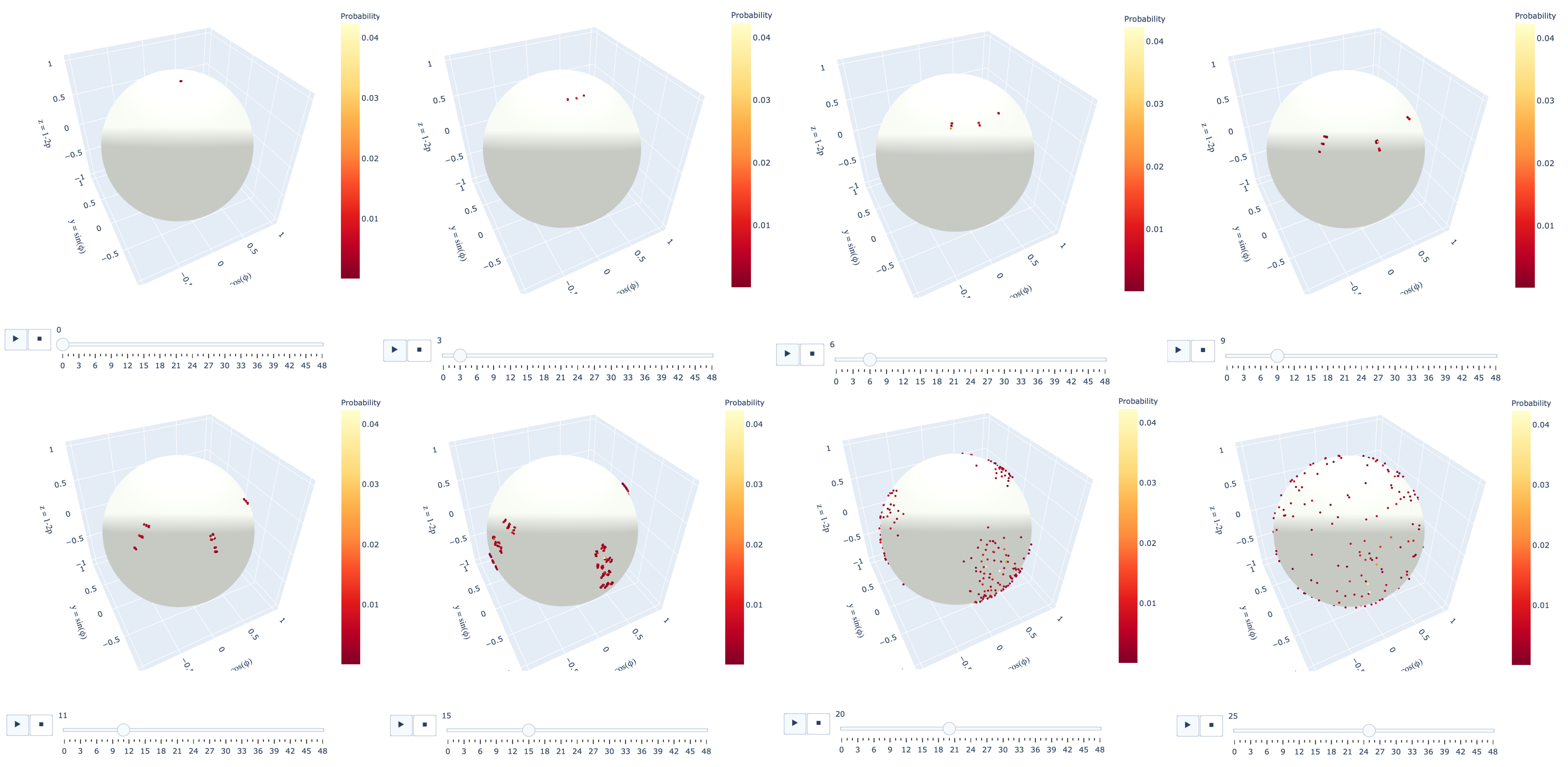}
\caption{Evolution of the geometric quantum state of one qubit interacting with 
	nine others via an Ising model with transverse field, visualized on the surface of the Bloch
	sphere. Each particle is $\Gamma_\alpha(t)$, represented in spherical coordinates 
	$\Gamma_\alpha(t) = \left( \theta_\alpha(t),\phi_\alpha(t)\right)$, with $\ket{\psi(\theta,\phi)} = 
	\cos \theta/2 \ket{0}+ \sin \theta/2 e^{i\phi}\ket{1}$. The color of the point encodes the 
	probability $x_\alpha(t)$. Time increases left to right and top to bottom. Each particle
	carries a probability mass $x_\alpha$, which is the probability to find the system in a 
	state $\Gamma_\alpha$. Thus, we think of it as an information carrier, moving the 
	information about the state of a quantum system around the quantum state space.
	The position of each particle is determined by the state of the environment and, indeed, 
	there are $2^9$ particles. The interactive html file from which these snapshots were
	taken can be found at \url{http://csc.ucdavis.edu/~cmg/papers/GeoStateEvolution.html}
	}
\label{fig:gqs_dynamics}
\end{figure*}
\paragraph*{Probability mass in a region of the state space.} Before we proceed with the
main result, we introduce here a notation that will be useful later. This is imported
from measure theory. Given a region $A \subseteq \PH$, we have
\begin{equation}
\mu_t(A) = \int_A q_t(Z) dV_{FS}
\end{equation}
$\mu_t(A)$ quantifies the probability that, at time $t$, our quantum system is in a state $Z$ that 
belongs to the region $A$ of the quantum state space $\PH$. In this sense, this is the core quantity
that conveys how much information about the system is contained in a certain region $A$
of the state space.

\section{Information Transport: Continuity equation}
\label{sec:IT}

In the past section we have summarized previous results about GQM. We now build on them 
and, by bringing in the dynamics of the system, we derive 
a continuity equation which dictates how the geometric quantum state of a non-equilibrium 
open quantum system evolves, under very general assumptions. This is the fundamental 
kinetic equation governing how the information about the state of a quantum system 
changes as a result of its interaction with an environment. 
Throughout this section we will try to maintain a fairly general language, to emphasize
how the treatment applies to quantum systems under very general assumptions. However, 
for concrete examples we will always refer to the simple case of a qubit.

\subsection*{General treatment}
The following treatment, and its results, pertains quantum systems which are finite-dimensional, 
and interact with finite-dimensional environments but are, otherwise, arbitrary. 
\begin{figure*}[t!]
\centering
\begin{minipage}[t]{.45\textwidth}
\includegraphics[width=\textwidth]{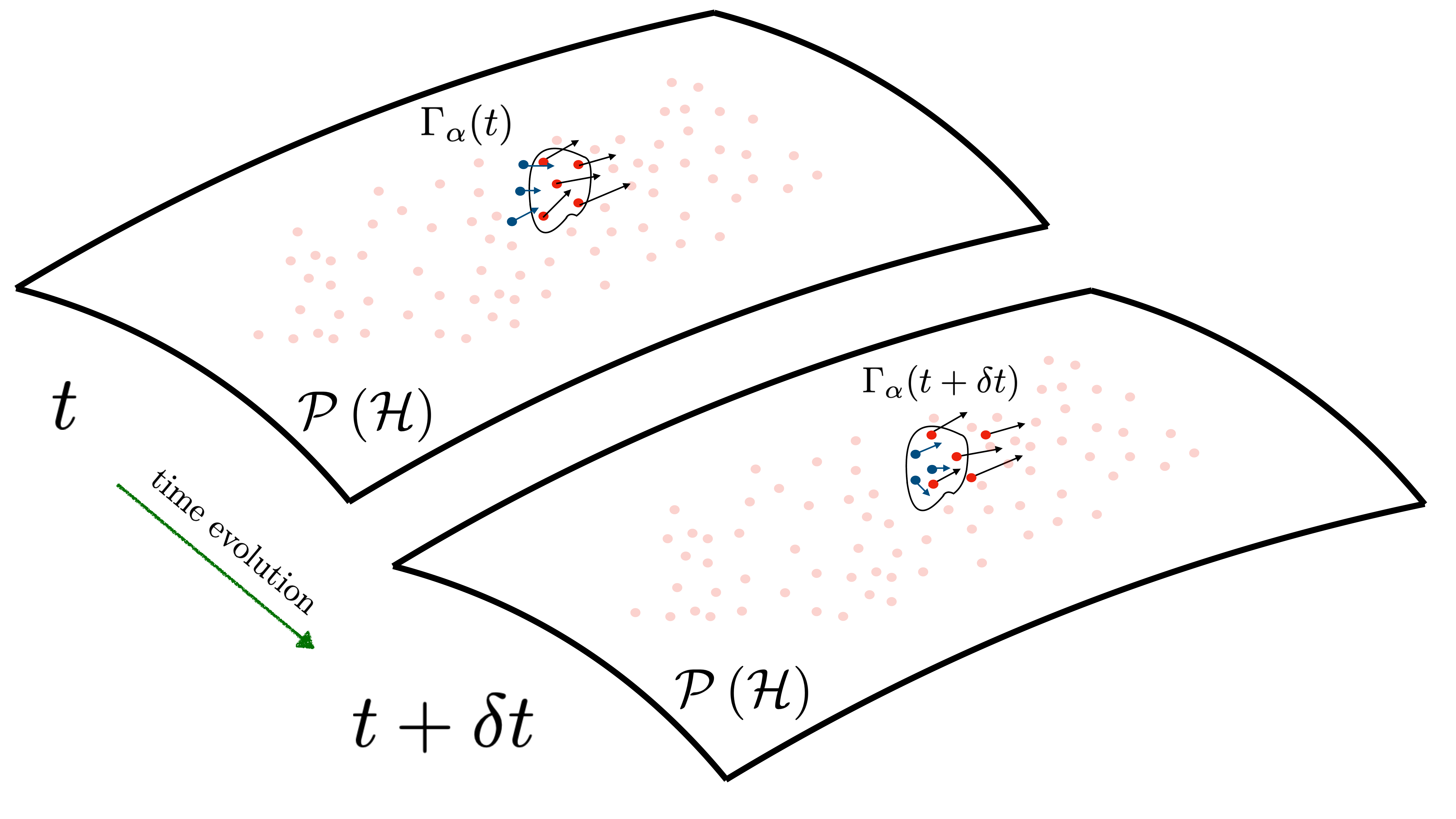}
\caption{Kinetic interpretation of the Information flux $J_t(Z)$. If we look at a small region of the 
	quantum state space,  due to the underlying dynamics we see that information is locally conserved 
	in the sense that there is a certain number of points which enters and leaves this region. As a 
	result of this local process, probability is moved around and can concentrate in a certain region 
	or get scrambled across the quantum state space.
	}
\label{fig:flux_term}
\end{minipage}\hfill
\begin{minipage}[t]{.45\textwidth}
\includegraphics[width=\textwidth]{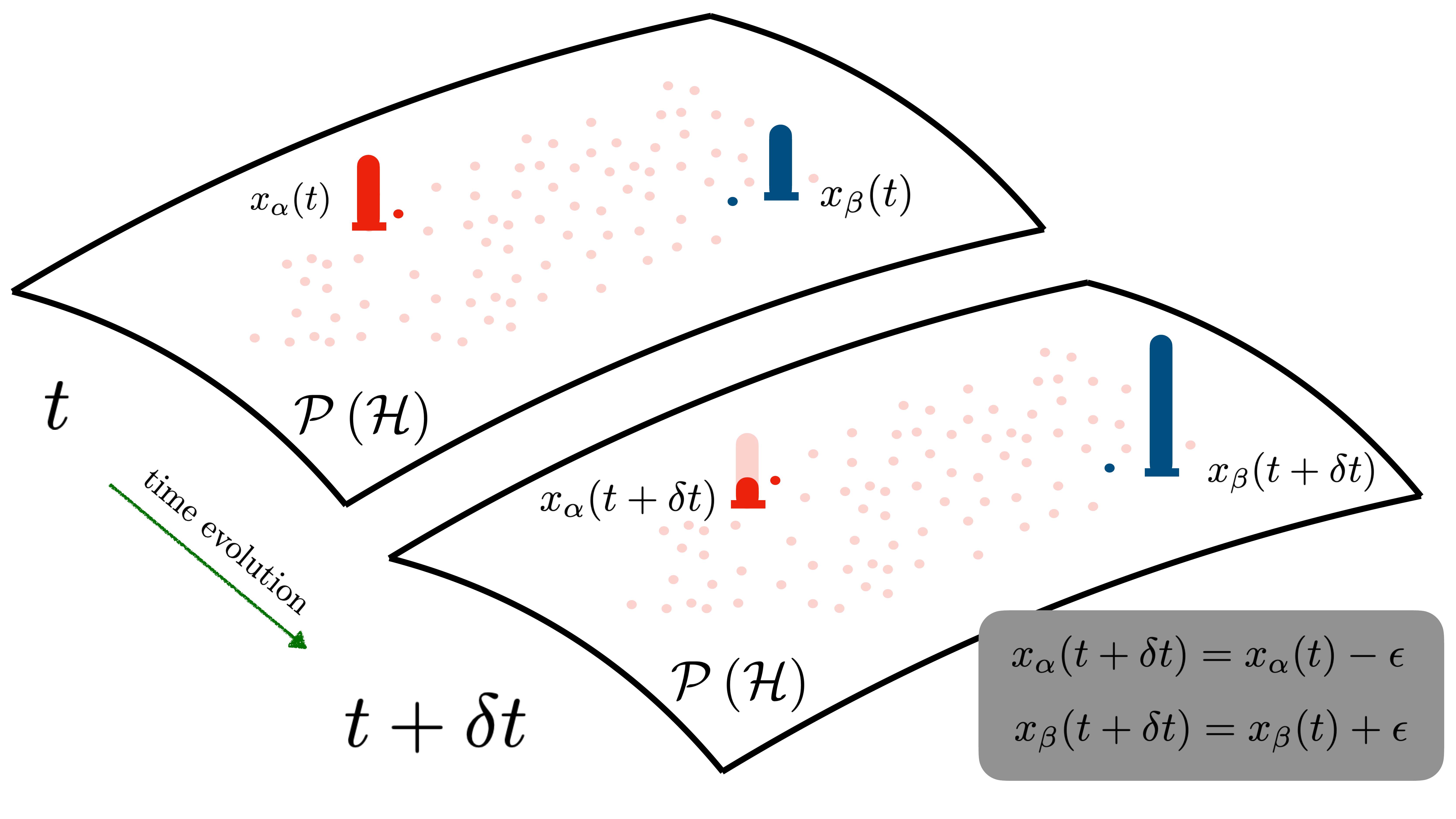}
\caption{Kinetic interpretation of the sink/source term $\sigma_t$. Even if the points do not move around
	the quantum state space $\dot{\Gamma}_\alpha=0$, the term $\sigma_t$ allows the exchange of information 
	between different regions. Note that this can be a non-local effect in state space, as depicted above. In the 
	example above	we have two states $\Gamma_\alpha$ and $\Gamma_\beta$ fixed in time, but exchanging 
	a certain amount $\epsilon$ of probability, thus moving information from one region of the state space to 
	another one: $x_\beta(t+\delta t) = x_\beta(t)+\epsilon$, $x_\alpha(t+\delta t) 	= x_\alpha(t) - \epsilon$.
	}
\label{fig:sigma_term}
\end{minipage}
\end{figure*}
A geometric quantum state is specified by two sets of quantities: $\left\{x_\alpha\right\}_{\alpha=0}^{d_E-1}$
and $\left\{ \Gamma_\alpha\right\}_{\alpha=0}^{d_E-1}$ (see Eq. (\ref{eq:xgammas})) which is a short-hand notation for 
$\Gamma_\alpha = Z(\ket{\chi_\alpha})$. The first one is a classical probability distribution
resulting from measuring the environment on a generic eigenbasis $\left\{\ket{b_\alpha} \right\}$. For each of the $x_\alpha$
there is a corresponding pure state that the system inhabits: $\Gamma_\alpha \in \mathcal{P}(\mathcal{H}_S)$,
which corresponds to the ket $\ket{\chi_\alpha} \in \mathcal{H}_S$. Since these $\Gamma_\alpha(t)$ are points moving 
in a state space, we can think of them as particles on a classical phases space, 
which interact in a non-trivial way. These are the ``carriers of information'', in the sense that each of 
these particles carries a probability mass $x_\alpha(t)$ that the system will be found in $\Gamma_\alpha(t)$.
A clarifying example of how the geometric quantum state of a qubit evolves when the system 
interacts with an environment is given in Figure \ref{fig:gqs_dynamics}. Since the total amount 
of information has to be preserved, $\mu_t(\PH) = \int_{\PH} q_t(Z) dV_{\mathrm{FS}} = \sum_\alpha x_\alpha(t) =1$,
the geometric quantum state $q_t(Z)$ must satisfy a continuity equation. Conceptually, this is 
the starting point of virtually all transport theories, which deal with quantities that are globally 
conserved, but that are moved around as a result of the underlying dynamics.

Thus, in the dynamics of a geometric quantum state we identify two different terms, arising from the 
time-evolution of these two sets of quantities. Loosely speaking, by looking at Figure \ref{fig:gqs_dynamics}
we can see that there are two different ways in which the geometric quantum state changes 
in time. First, there is the color of each particle, which changes over time: $\dot{x}_\alpha \neq 0$. Second, there is
the movements of each point: $\dot{\Gamma}_\alpha \neq 0$. By summing each term over
all particles, and then summing the two terms, we get the following evolution equation for the geometric quantum state:
\begin{equation}
\frac{\partial q_t(Z)}{\partial t} = \sum_\alpha \dot{x}_\alpha \tilde{\delta}\left( Z- \Gamma_\alpha \right) - x_\alpha \widetilde{\delta}^{'}\left(Z - \Gamma_\alpha \right) \dot{\Gamma}_{\alpha}~.\label{eq:cont}
\end{equation}
Let's start by analyzing the second term. There, $\widetilde{\delta}^{'}$ is the first distributional derivative 
of $\widetilde{\delta}(Z-\Gamma_\alpha)$ and $\Gamma_\alpha(t)$ is the velocity of each particle $\Gamma_\alpha(t)$.
Thus, the second term in the right-hand side looks like the divergence of a velocity field or, in other words, as a flux
term in a continuity equation. Indeed, since all the dependence on the coordinate $Z$ lies inside the $\widetilde{\delta}^{'}$, 
the whole term can be written as a covariant divergence\footnote{With an arbitrary coordinate system we have 
$\mathrm{div} \vec{f} = \frac{1}{\sqrt{g}} \partial_\alpha \left(\sqrt{g} f^\alpha\right)$,
where $g = \vert \mathrm{det} g_{FS}\vert$ is the absolute value of the determinant of the metric which, 
in our case, is the Fubini-Study metric.}
\begin{equation}
\sum_\alpha x_\alpha \widetilde{\delta}^{'}\left(Z - \Gamma_\alpha \right) \dot{\Gamma}_{\alpha} = (\nabla \cdot J_t)(Z)~,
\end{equation}
where we have identified the \emph{Information Flux} $J_t(Z)$ as the ``single-particle'' flux $J_\alpha(Z,t)$, averaged
with the probability mass it is carrying $x_\alpha(t)$:
\begin{subequations}\label{eq:flux}
\begin{align}
&J_\alpha(Z,t) =  \widetilde{\delta}\left(Z - \Gamma_\alpha(t) \right) \dot{\Gamma}_{\alpha}(t)~,\\
&J_t(Z) = \sum_\alpha x_\alpha(t)J_\alpha(Z,t)~.
\end{align}
\end{subequations}
We now look at the first term in the right-hand side of Eq.(\ref{eq:cont}): $\sum_\alpha \dot{x}_\alpha(t) \delta(Z-\Gamma_\alpha(t))$.
We recognize a source/sink term, which is independent on the underlying movement of
the points in the quantum state space. This identifies the other functional term in our continuity 
equation: \emph{sinks and sources of information} $\sigma_t(Z)$:
\begin{equation}
\sigma_t(Z) = \sum_{\alpha=0}^{d_E} \dot{x}_\alpha(t) \widetilde{\delta}\left(Z - \Gamma_\alpha(t) \right)~.\label{eq:sigma}
\end{equation}
Eventually, we obtain the following continuity equation for the geometric quantum state of a finite-dimensional
quantum system interacting with a finite-dimensional environment:
\begin{equation}
\frac{\partial q_t}{\partial t} + \nabla \cdot J_t = \sigma_t\label{eq:continuity}
\end{equation}
This is our first result. A crucial aspect here is that this theoretical framework allows us to talk about information 
as a physically localized (and overall conserved) quantity, which is carried around by points in quantum state space.
This is in analogy with most classical theories of transport which deal with properties of particles, such as mass 
or charge, which are carried around by particles, represented as points wondering in a classical phase-space.

To support this theoretical framework, the physical interpretation of the flux $J_t$ and source/sink $\sigma_t$ term is essentially the 
same as in other transport theories. We now describe them explicitly, using their kinetic interpretation, drawn
in Figures \ref{fig:flux_term} and \ref{fig:sigma_term}. First, $J_t$ is an information flux, in the proper sense of flux. 
This flux is associated to an overall conserved quantity, probability mass, localized in its carriers $\Gamma_\alpha$, which 
is moved around the quantum state space. To emphasize this point, we look at the simpler situation in which $\dot{x}_\alpha = 0$. 
In this case, $\sigma_t=0$ and the information carried around by the points in state space is a constant of motion 
($x_\alpha(t)= x_\alpha(t_0)$). Thus, the analogy with a model of point particles carrying around a physical property, 
like charge or mass, becomes exact. As the carriers move around, the information is dispersed across the quantum 
state space with a mechanisms that ensures it is locally preserved, and therefore well described with a continuity 
equation that equates the local time derivative of the probability distribution with the divergence of the local flux of 
particles.

Second, the term $\sigma_t$ is a proper source/sink term, with the usual interpretation. Indeed, first, we note that it
can not be written as a divergence term. Second, we look at the simpler dynamics in which the position of the points 
in $\PH$ does not change: $\dot{\Gamma}_\alpha = 0$. In this case the support of the distribution is fixed and the 
particles don't move around $\Gamma_\alpha(t)=\Gamma_\alpha(t_0)$. However, there can still be a non-trivial 
dynamics, due to $\dot{x}_\alpha(t)\neq 0$. Depending on the actual support of the distribution, and how spread it is, 
as depicted in Figure \ref{fig:sigma_term}, the time-evolution generated by this term is generally non-local in the 
quantum state space. Thus, $\sigma_t$ possesses all the hallmarks of the standard sinks and sources terms in 
transport theories: it is not associated with particles moving around the state space and it can not be rewritten 
as a divergence term.

<<<<<<< HEAD
\subsection{Isolated quantum systems}\label{subsec:IQS}
=======
\subsection*{Isolated quantum systems}
>>>>>>> 4244be5094661ebf870f12252a225135918b1013

What happens when the system is closed? Is there a specific form for the information flux $J_t$ and source $\sigma_t$?
Here we answer both these questions in a detailed manner. Since the evolution is Hamiltonian, we can 
use Hamilton's equations of motion to derive the evolution equation of $\mu_t$ or, equivalently, of $q_t(Z)$.
First, the sink/sources terms vanishes, since the evolution is Hamiltonian, thus preserving the local probabilities.
Thus proves that a $\sigma_t \neq 0$ is related to the existence of a dissipative dynamics. Second, starting from 
the definition of the flux in Eq.(\ref{eq:flux}), we can explicitly write the form of the flux $J_t$, 
using Hamilton's equations of motion. Writing $\Gamma_\alpha$ in canonical coordinates $(p_n(\Gamma_\alpha),\phi_n(\Gamma_\alpha))$, 
we have 
\begin{align}
\dot{\Gamma}_\alpha & = \left( \dot{p}_n(\Gamma_\alpha), \dot{\phi}_n(\Gamma_\alpha)\right)_n \nonumber \\
& = \left( \frac{1}{\hbar}\frac{\partial E}{\partial \phi_n}, -\frac{1}{\hbar}\frac{\partial E}{\partial p_n}  \right)_n = v_H(p_n,\phi_n)
\end{align}
Note how the right-hand side, which we called $v_H$ does not depend on the index $\alpha$ anymore: all 
the points evolve in the same way, following the same Hamiltonian flow. Inserting this into the definition of the flux 
we obtain that the flux is equal to the product between the distribution and the Hamiltonian velocity field $v_H$:
\begin{equation}
J_t(Z) = q_t(Z) v_H(Z)~,
\end{equation}
thus providing the following continuity equation:
\begin{equation}
\frac{\partial q_t}{\partial t} = - v_H \cdot \nabla q_t - q_t \, \nabla \cdot v_H
\end{equation}
This can be further simplified by noting that the Hamiltonian vector field is divergence free, $\nabla \cdot v_H = 0$.
This is due to the smoothness of $E(p_n,\phi_n)$ which, via Schwarz's theorem, implies that the Hessian of $E(p_n,\phi_n)$ 
is symmetric. 
\begin{align}
\nabla \cdot v_H &= \sum_n \left( \frac{\partial}{\partial p_n} , \frac{\partial}{\partial \phi_n}\right) \cdot \frac{1}{\hbar}\left(\frac{\p E}{\p \phi_n}, -\frac{\p E}{\p p_n} \right) \nonumber \\
& = \frac{1}{\hbar} \sum_n \left( \frac{\p^2 E}{\p p_n \p \phi_n } - \frac{\p^2 E}{\p \phi_n \p p_n }\right) = 0 \nonumber
\end{align}

This leads us to the final form of the continuity equation of an isolated quantum system:
\begin{equation}
\frac{\partial q_t}{\partial t} = - v_H \cdot \nabla q_t = -\left\{ q_t, E \right\}
\end{equation}

\paragraph*{Liouville's theorem for GQM.} As a further point of contact with the techniques of classical 
statistical mechanics and kinetic theory, we now show that a generic Hamiltonian dynamics for the 
geometric quantum state satisfies Liouville's theorem \cite{Soto16}. Indeed, by writing explicitly the total 
derivative of $q_t(Z)$ with respect to time we get 
\begin{equation}
\frac{d q}{d t} = \sum_\alpha \frac{\partial q_t}{\partial p_\alpha} \frac{dp_\alpha}{dt}+\frac{\partial q_t}{\partial \phi_\alpha} \frac{d\phi_\alpha}{dt} + \frac{\partial q_t}{\partial t}
\end{equation}
Inserting Hamilton's equations of motion (Eq. (\ref{eq:HAM_EOM})), and then using the continuity equation
\begin{align}
\frac{d q}{d t} &= \sum_\alpha \frac{\partial q_t}{\partial p_\alpha} \frac{dp_\alpha}{dt}+\frac{\partial q_t}{\partial \phi_\alpha} \frac{d\phi_\alpha}{dt} + \frac{\partial q_t}{\partial t}\nonumber \\
& = \left\{ q_t, E\right\} - \nabla \cdot q_t v_H = 0
\end{align}
This concludes the section about the general approach, and the treatment of isolated quantum systems.
We now turn to the more convoluted case of an open quantum system.

\section{Kinetic equations of information transport in open quantum systems}
\label{sec:DYN}

As we move to analyze how open quantum systems scramble information around the
quantum state space, the goal of this section is to provide a microscopic approach
to the transport of information: a kinetic theory of how the information about the state
of a quantum system changes as a result of its interaction with a structured, non-thermal,
environment. Thus, the main outcome of this section is a concrete set of microscopic 
equations that determines the evolution of the geometric quantum state in a non-Hamiltonian 
setting.

Calling $H_S$ and $H_E$ the Hamiltonian operators of the system and environment, respectively, 
the total Hamiltonian of the joint system is $H = H_S + H_E + H_{\mathrm{int}}$. Since $H_{\mathrm{int}}$
is the interaction term between system and environment, we can always put it in the form
\begin{equation}
H_{\mathrm{int}} = \sum_{k=1}^M A^{(k)} \otimes B^{(k)}~,
\end{equation}
where $A^{(k)}$ and $B^{(k)}$ are operators with support on $\mathcal{H}_S$ and $\mathcal{H}_E$,
respectively. With a slight abuse of notation we will often conflate $A^{(k)}$ with $A^{(k)}\otimes \mathbb{I}_E$
and $B^{(k)}$ with $\mathbb{I}_S \otimes B^{(k)}$, where $\mathbb{I}_S$ and $\mathbb{I}_E$ are, respectively,
the identity operator on $\mathcal{H}_{S}$ and $\mathcal{H}_E$.
Here we do not impose a specific choice for the basis of the system and environment. While a natural one
is to choose the bases that diagonalize the non-interacting Hamiltonians $H_S$ and $H_E$, there are cases
where a different choice is more appropriate. For example, in the case of a spin-$1/2$ chain one might be interested 
in using the computational basis. Thus, we keep things general and use a generic basis, with no particular 
properties with respect to the algebra of observables: $\left\{ \ket{a_j}\right\}_{j=0}^{d_S}$ and $\left\{ \ket{b_\alpha}\right\}_{\alpha=0}^{d_E-1}$.

Since the goal is to derive a dynamic equation for $\left\{x_\alpha(t) \right\}$ and $\left\{\Gamma_\alpha(t)\right\}$,
we begin with the overall Schroedinger equation for the pure state $\ket{\psi(t)}$ of the joint system+environment, written
in the generic tensor product basis defined above, with $\psi_{j\alpha}(t) = \braket{a_j,e_\alpha}{\psi(t)}$:
\begin{equation}
i\hbar \frac{d \psi_{j\alpha}(t)}{dt} = \sum_{k,\beta} H_{j\alpha;k\beta} \psi_{k\beta}(t)~,
\end{equation}
where
\begin{align}
H_{j\alpha;k\beta} &\coloneqq \bra{a_j,b_\alpha} H \ket{a_k,b_\beta}\\
&= \left(H_S\right)_{jk}\delta_{\alpha \beta} + \delta_{jk} \left(H_E\right)_{\alpha \beta} + \sum_{n=1}^M A_{jk}^{(n)} B^{(n)}_{\alpha \beta}~.\nonumber 
\end{align}
We now collect the probability $x_\alpha(t)$ and the states $\ket{\chi_\alpha(t)}$ into
a single quantity: a non-normalized ket $\ket{\Phi_\alpha(t)} \coloneqq \sqrt{x_\alpha}\ket{\chi_\alpha} \in \mathcal{H}_S$.
By plugging Schroedinger's equation into the time-derivative of $\ket{\Phi_\alpha(t)}$ we get the following set of $d_E$
coupled linear equations:
\begin{equation}
i\hbar \frac{d \ket{\Phi_\alpha}}{dt} = H_S \ket{\Phi_\alpha} + \sum_{\beta} \left( H_E\right)_{\alpha\beta}\ket{\Phi_\beta} + \hat{M}_{\alpha \beta}\ket{\Phi_\beta}~,
\end{equation}
where $\hat{M}_{\alpha\beta}$ is a set of operators acting on the system, defined as 
\begin{equation}
\hat{M}_{\alpha \beta} \coloneqq \sum_{k=1}^M B^{(k)}_{\alpha\beta} A^{(k)}
\end{equation}
This can be further manipulated to separate the non-interacting part, acting on each vector $\ket{\Phi_\alpha}$, 
from the interacting part, acting on $\ket{\Phi_\beta}\neq \ket{\Phi_\alpha}$. The final form of our kinetic equation is
\begin{equation}
i\hbar \frac{d \ket{\Phi_\alpha}}{dt} = \hat{H}_\alpha \ket{\Phi_\alpha} + \sum_{\beta \neq \alpha} \hat{V}_{\alpha \beta}\ket{\Phi_\beta}~,\label{eq:kinetic}
\end{equation}
where the single-particle Hamiltonian is $\hat{H}_\alpha$ is
\begin{equation}
\hat{H}_\alpha \coloneqq H_S + \left( H_E\right)_{\alpha \alpha} \mathbb{I}_S + \sum_{k=1}^M B^{(k)}_{\alpha \alpha}A^{(k)}~,\label{ed:def_Halpha}
\end{equation}
and the interaction between particles is vehiculated by the set of operators $\hat{V}_{\alpha \beta}$
\begin{equation}
\hat{V}_{\alpha \beta} \coloneqq \left( H_E\right)_{\alpha \beta} \mathbb{I}_S + \sum_{k=1}^M B^{(k)}_{\alpha \beta} A^{(k)}~.\label{ed:def_V}
\end{equation}
The remaining step is to connect this with the quantities that determine the geometric quantum state of the system: $x_\alpha$ and $\Gamma_\alpha$.
This is easily done by using the original definition $\ket{\Phi_\alpha} = \sqrt{x_\alpha} \ket{\chi_\alpha}$. Since $\Gamma_\alpha(t) \leftrightarrow \ket{\chi_\alpha}$,
we have $\Gamma^j_\alpha(t) = \frac{\braket{a_j}{\Phi_\alpha(t)}}{\sqrt{x_\alpha(t)}}$ and $x_\alpha(t) = \braket{\Phi_\alpha(t)}{\Phi_\alpha(t)}$.


Equation (\ref{eq:kinetic}) provides a way to describe the dynamics of an open 
quantum system as a set of $d_E$ linear, coupled, differential equations, for the 
non-normalized states $\ket{\Phi_\alpha}$. The emergence of this set of coupled 
equation also reinforces the interpretation of $\Gamma_\alpha(t)$ as behaving as 
``particle-like'', with their own local notion of energy, represented by $H_\alpha$ and 
their pair-wise interaction $\hat{V}_{\alpha \beta}$. While not treated here, by using 
the approach developed in Ref.\cite{Anza20a} these results can be extended 
to include finite-systems interacting with infinite-dimensional environments.

As a consistency check, we note that the preservation of information within the 
quantum state space of the system is inherited by the unitarity of the global evolution. 
This is manifested in the relations $\hat{H}_\alpha^\dagger = \hat{H}_\alpha$ and 
$\hat{V}_{\alpha \beta}^{\dagger} = \hat{V}_{\beta \alpha}$, both of which can be 
easily shown to be true from the definitions above.

So far, we have derived the microscopic, kinetic, equations regulating the 
information transport across the quantum state space. In the next few sections, using
both analytical and numerical approaches, we explore the theoretical framework developed
so far to look at the phenomenology of information transport in concrete physical systems. 
In particular, in our examples we look at the information transport in the state 
space of a qubit, both in isolated and open configurations. This choice is made 
to support the theoretical calculations with visual representations, both of which 
are simpler in the case of a qubit.

\section{System 1: Isolated qubit}
\label{sec:EXAMPLES1}

\begin{figure*}[t!]
\centering
\includegraphics[width=.9\textwidth]{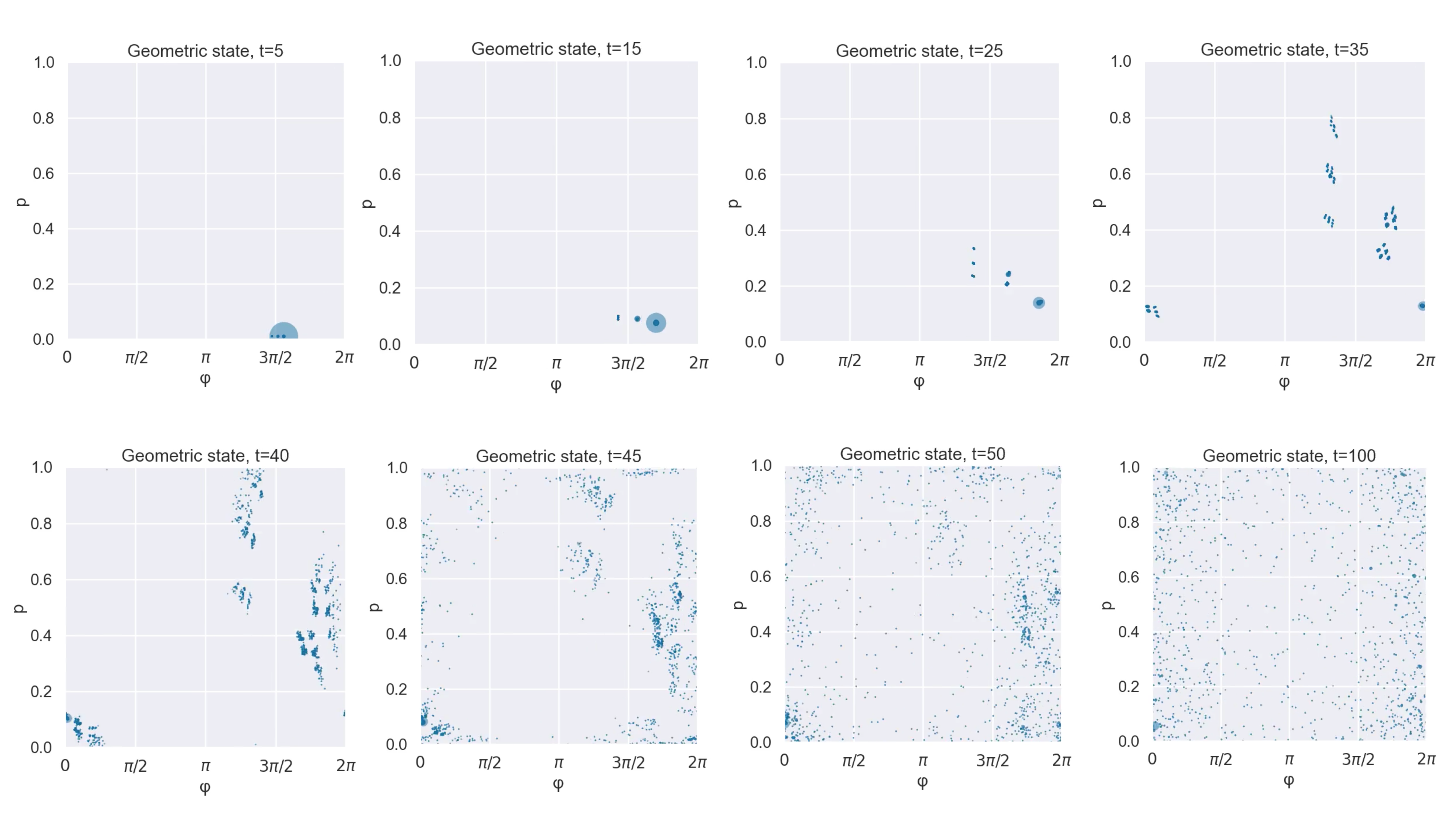}
\caption{Evolution of the geometric quantum state of one qubit interacting with 
	eleven others via an antiferromagnetic Ising model with both transverse and longitudinal field $\vec{B} = (1,0,0.5)$, 
	visualized on the Bloch square. Each particle is $\Gamma_\alpha(t)$, represented in canonically conjugated 
	coordinates $\Gamma_\alpha(t) = \left( p=p_\alpha(t),\phi=\phi_\alpha(t)\right)$, with $\ket{\psi(p,\phi)} = 
	\sqrt{1-p} \ket{0}+ \sqrt{p} e^{i\phi}\ket{1}$. The radius of the particle encodes its 
	probability mass $x_\alpha(t)$. Time increases left to right and top to bottom. Each particle
	carries a probability mass $x_\alpha$, which is the probability to find the system in a 
	state $\Gamma_\alpha$. Thus, we think of it as an information carrier, moving the 
	information about the state of a quantum system around the quantum state space.
	After $60/70$ time steps the distribution settles on an equilibrium distribution. A full video 
	of the dynamical evolution is available at the following link: \url{https://github.com/fabioanza/GeomQuantMech/blob/main/video_g1_h05_L12.mp4}
	}
\label{fig:gqs_dynamics2}
\end{figure*}

Before we begin our case studies, here we collect useful technical details
about the geometric quantum mechanics of a qubit. Then, using as a reference 
basis an arbitrary basis $\left\{ \ket{0},\ket{1}\right\}$ 
the (probability,phase) coordinates are identified via the scalar product operation
on the Hilbert space $\braket{s}{\psi} = \sqrt{p_s}e^{i\phi_s}$, with $s=0,1$, giving $(p_0,\phi_0,p_1,\phi_1)$. 
However, the quantum state space gets rid of two fundamental redundancies in this description.
First, we can always assume that $\phi_0 = 0$. Second, due to normalization we have
$p_0 = 1-p_1$. Thus, we only have a pair of independent coordinates: $(p_1,\phi_1)$.
We can therefore drop the index and say that a general point on the state space
of a qubit is uniquely identified by $(p,\phi) \in [0,1]\times[0,2\pi[$, with an embedding on the Hilbert space
defined by $\ket{\psi(p,\phi)} = \sqrt{1-p}\ket{0} + \sqrt{p}e^{i\phi}\ket{1}$. In analogy with 
the Bloch sphere, we call the square representation of $\mathbb{C}P^1$ the ``Bloch square''.
Note that the embedding is smooth everywhere, aside for two isolated points $\ket{0}$ and 
$\ket{1}$, where $\phi$ is not defined and which are represented only by $p=0$ and $p=1$, 
respectively. These coordinates $(p,\phi)$ are canonically conjugated: $\left\{ p,\phi\right\}$
and the determinant of the Fubini-Study metric is simply $g= \frac{1}{2}$. Moreover, since the total
volume of the state space is $\pi$ \cite{Bengtsson2017}, we renormalize the Fubini-Study volume element 
with the total volume, so that $dV = \frac{dpd\phi}{2\pi} = \frac{dV_{FS}}{\mathrm{Vol}(\mathbb{C}P^1)}$.

\subsection*{Analytical solution}

Our first case study is an isolated qubit, with an arbitrary initial geometric quantum 
state $f_0(p,\phi)$ and Hamiltonian operator $H = \vec{B} \cdot \vec{m} = E_0 \ket{E_0}\bra{E_0} + E_1 \ket{E_1}\bra{E_1}$,
with $\vec{B}$ an arbitrary $3D$ magnetic vector field, $\vec{m} = \frac{\hbar}{2}\vec{\sigma}$
and $\vec{\sigma}$ are the Pauli matrices. Thus, the generator of the geometric evolution is:
\begin{align}
\mathcal{E}(p,\phi) & = \bra{\psi(p,\phi)}H\ket{\psi(p,\phi)} = (1-p)H_{00} + p H_{11} \nonumber \\
&+ \sqrt{p(1-p)} \left(e^{i\phi}H_{01}+e^{-i\phi}H_{10} \right) \\
& = \vec{B} \cdot \vec{S} \nonumber~,
\end{align}
where 
\begin{equation}
\vec{S}= \MV{\vec{m}} = \frac{\hbar}{2}\left( 2\sqrt{p(1-p)}\cos \phi, 2\sqrt{p(1-p)}\sin \phi , p\right)\nonumber
\end{equation}

\begin{figure*}[t!]
\centering
\includegraphics[width=.9\textwidth]{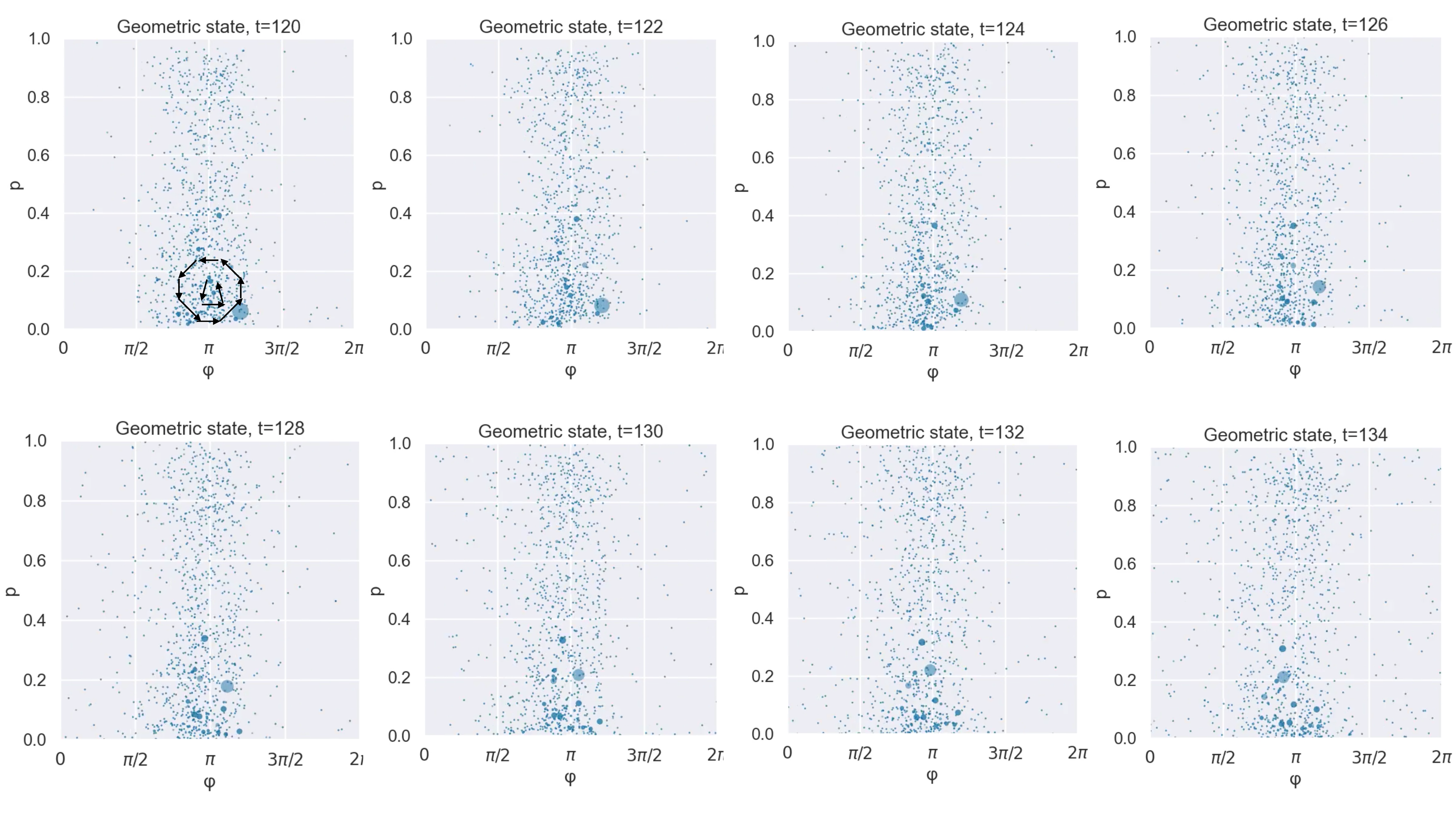}
\caption{Evolution of the geometric quantum state of one qubit interacting with 
	eleven others via a Ferromagnetic Ising model with longitudinal and transverse field, 
	visualized on the Bloch square. Each particle is $\Gamma_\alpha(t)$, represented via
	canonically conjugated coordinates $\Gamma_\alpha(t) = \left( p=p_\alpha(t),\phi=\phi_\alpha(t)\right)$, with $\ket{\psi(p,\phi)} = 
	\sqrt{1-p} \ket{0}+ \sqrt{p} e^{i\phi}\ket{1}$. The radius of the particle encodes its 
	probability mass $x_\alpha(t)$. Time increases left to right and top to bottom. Each particle
	carries a probability mass $x_\alpha$, which is the probability to find the system in a 
	state $\Gamma_\alpha$. Here we show the detail of the evolution of a macroscopic 
	coherent phenomenon that gives rise to sustained fluctuations and maintains the
	system out of thermal equilibrium. A full video of the dynamical evolution is available 
	\url{https://github.com/fabioanza/GeomQuantMech/blob/main/video_g1_h05_L12_ferro_long.mp4}
	}
\label{fig:gqs_dynamics3}
\end{figure*}
Since the evolution is Hamiltonian, there is no source or sink term, so the continuity equation is 
\begin{equation}
\frac{\partial f_t(p,\phi)}{\partial t} + \nabla \cdot J_t = 0~.
\end{equation}
In canonical coordinates, the $\nabla$ operator is simply $\nabla = (\frac{\p }{\p p},\frac{\p }{\p \phi})$.
Writing explicitly, we get
\begin{equation}
\frac{\partial f_t(p,\phi)}{\partial t} + \frac{\p J_t^p}{\p p} + \frac{\p J_t^{\phi}}{\p \phi} = 0~.
\end{equation}
We now come to writing explicitly the flux $J_t = (J_t^p,J_t^{\phi})$. As argued above, 
with Hamiltonian evolution the flux is the product between the Hamiltonian vector 
field $v_H$ and the geometric quantum state $f_t(p,\phi)$. Explicitly:
\begin{equation}
J_t =  \frac{f_t(p,\phi)}{\hbar} \left(\frac{\p \mathcal{E}(p,\phi)}{\p \phi}, -\frac{\p \mathcal{E}(p,\phi)}{\p p}\right)
\end{equation}
Plugging this into the continuity equation and remembering that a Hamiltonian vector field is 
divergence-free, we the following equation for $f_t(p,\phi)$
\begin{equation}
\frac{\partial f_t}{\partial t} + A(p,\phi) \frac{\partial f_t}{\partial p} + B(p,\phi) \frac{\partial f_t}{\partial \phi} = 0~,
\end{equation}
with $A(p,\phi) = \frac{1}{\hbar}\frac{\partial \mathcal{E}}{\partial \phi}$ and $B(p,\phi) = - \frac{1}{\hbar}\frac{\partial \mathcal{E}}{\partial p}$.
Alternatively, using the poisson brackets, we get
\begin{equation}
\frac{\partial f_t}{\partial t} + \left\{ f_t, \mathcal{E}\right\} = 0
\end{equation}
Its solution is given by the propagator $P(t,t_0) = \mathrm{Exp}\left[ - (t-t_0)\left\{\mathcal{E}, \cdot \right\} \right]$.
Since $\mathcal{E}= \vec{h} \cdot \vec{m}$ and, by using the algebra of poisson brackets,
one can prove that $\vec{m} = \frac{\hbar}{2}\left( S_x(p,\phi),S_y(p,\phi),S_z(p,\phi)\right)$ are the generators of $SO(3)$:
\begin{equation}
\left\{ m_a, m_b\right\} = \epsilon_{abc} m_c~,
\end{equation}
 here we recognize that the propagator $P(t,t_0)$ implements a 3D rotation of the generic 
expectation value $\vec{m}(t)$ around the axis $\vec{B}$, with angle $||\vec{B}||(t - t_0)$. 
In more physics-agnostic terms, this is a linear, first-order, partial differential equation, which can be solved 
in full generality by using the method of the characteristics. Here we simplify the treatment 
by switching to coordinates aligned to the convenient basis in which the Hamiltonian 
is diagonal. This leads to
\begin{equation}
\frac{\partial f_t}{\partial t} - \omega \frac{\partial f_t}{\partial \phi} = 0~,
\end{equation}
where $\omega = \frac{E_1 - E_0}{\hbar}$. This is equivalent to the transport 
equation for advection, or heat equation, which leads to the analytical solution for the 
time-dependent geometric quantum state: $f_t(p,\phi) = f_0(p,\phi+\omega t)$, for any 
valid initial distribution $f_0(p,\phi)$.

\subsection*{Discussion}
\begin{figure*}[t!]
\centering
\includegraphics[width=.9\textwidth]{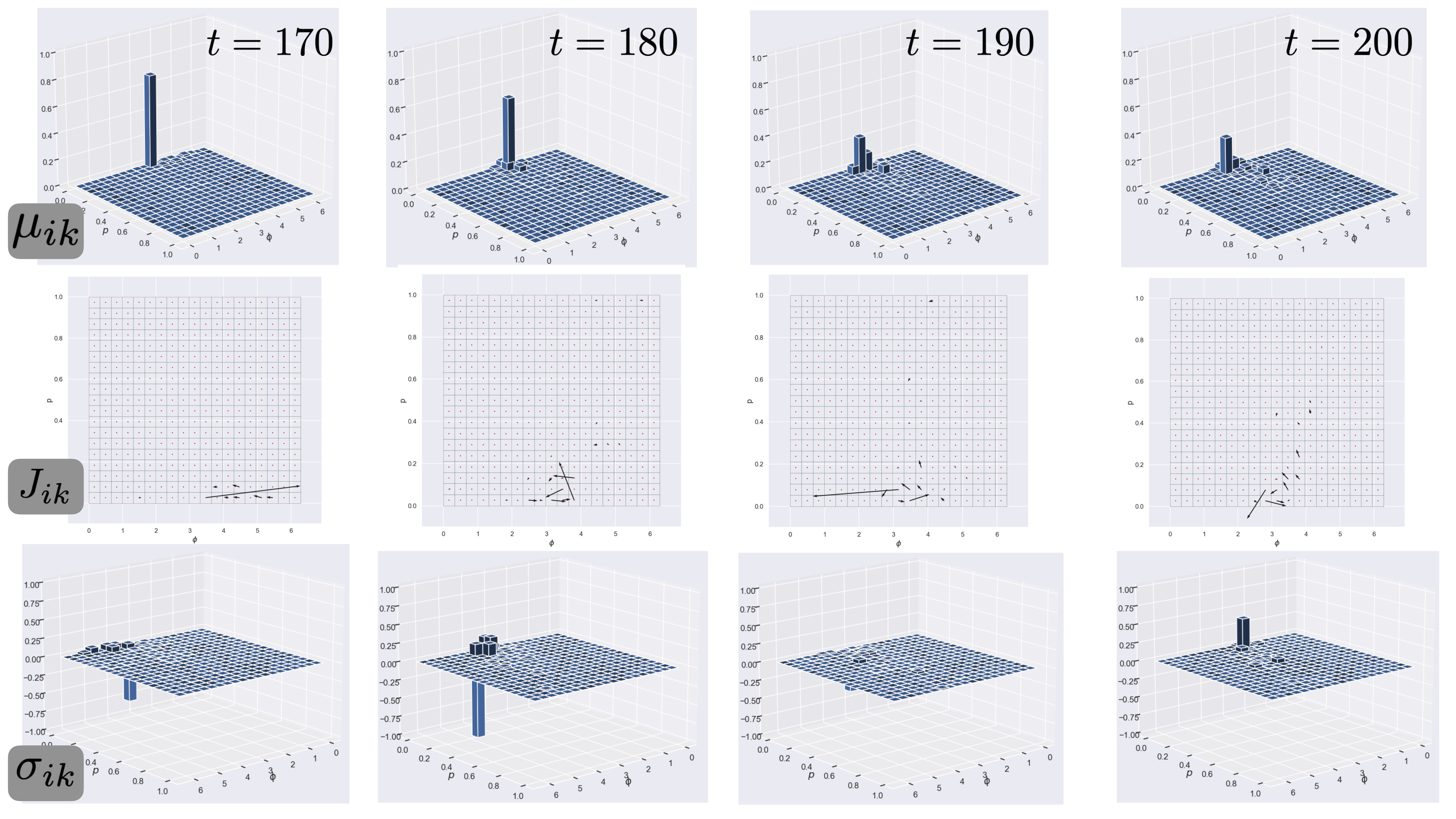}
\caption{Evolution of the coarse-grained geometric quantum state (Top row), Information Flux (Middle row)
	and Information sink/source (Bottom row), for one qubit interacting with eleven other ones via
	a Ferromagnetig Ising mode with longitudinal and transverse field, visualized on the Bloch square. 
	We have sliced the quantum state space in $(p,\phi)$ coordinates into a grid of $19 \times 19$ small 
	uniform cells $\mathcal{C}_{ik} = \left[\frac{i}{20},\frac{i+1}{20}\right[ \times \left[2\pi\frac{k}{20},2\pi \frac{k+1}{20}\right[$ of 
	equal volume and integrated the relevant quantities for the study of the continuity equation over $\mathcal{C}_{ik}$ 
	to obtain a time-dependent coarse-grained graphic representation: $\mu_{ik}(t)$, $J{ik}(t)$ and $\sigma_{ik}(t)$.
	}
\label{fig:summary}
\end{figure*}
The evolution of a isolated qubit is periodic in time and this appears in the evolution 
of its geometric quantum state as $f_0(p,\phi+\omega t)$. Crucially, as the 
distribution changes over time with a Hamiltonian evolution, it does so in a 
rigid fashion, never changing its shape. This can, indeed, be seen explicitly. 
Say that, at $t=0$, the geometric quantum state is made by a convex sum
of dirac deltas concentrated on a certain number $N$ of points $\Gamma_\alpha = (p_\alpha,\phi_\alpha)$: 
$q_0(p,\phi) = \sum_{\alpha=1}^N \lambda_\alpha 2\delta\left[ p-p_\alpha\right] \delta\left[ \phi-\phi_\alpha\right]$.
The evolution of the geometric quantum state is therefore dictated by how 
the points $\Gamma_\alpha(t)$ change over time. Since, however, the evolution 
can be written as a linear change in the $\phi$ coordinate $\phi(t)=\phi(0)+\omega t$, this translates
into $\phi_\alpha(t) = \phi_\alpha(0)- \omega t$, with $p_\alpha(t)=p_\alpha(0)$.
As a result, the shape of the distribution never changes as the relative distance
between the points $\Gamma_\alpha(t)$ does not depend on time. Calling $d_{FS}(\cdot,\cdot)$
the Fubini-Study distance, we have
\begin{align}
&\cos d_{FS}(p_\alpha, \phi_\alpha;p_\beta,\phi_\beta) \coloneqq \Big(1-(p_\alpha+p_\beta)+p_\alpha p_\beta \nonumber \\
&\left. +\sqrt{p_\alpha(1-p_\alpha)}\sqrt{p_\beta(1-p_\beta)}\cos (\phi_\alpha-\phi_\beta)\right)^{1/2},
\end{align}
giving
\begin{equation}
d_{FS}(\Gamma_\alpha(t),\Gamma_\beta(t))=d_{FS}(\Gamma_\alpha(0),\Gamma_\beta(0)) \quad \forall \alpha,\beta~.\label{eq:nochaos}
\end{equation}
Since Hamiltonian evolutions can generically support folding and stretching 
in state space, causing chaotic behavior, we conclude that this is not the 
most generic Hamiltonian evolution a qubit state space can support. Note
that this goes beyond the assumption of a time-independent Hamiltonian.
Even with time-dependent Hamiltonian, this aspect of the isolated evolution
of a qubit still holds true, thus confirming that a unitary evolution might not be
the most general Hamiltonian evolution that can occur in a quantum state space.

\section{System 2: Qubit in Ising environment}
\label{sec:EXAMPLES2}

We now look at the case of a single qubit which interacts with an environment 
made by a $1D$ chain of qubits evolving with an Ising Hamiltonian with periodic
boundary conditions $\vec{\sigma}_{k+L}= \vec{\sigma}_k$
\begin{equation}
H_{\mathrm{Ising}} = \sum_{k=0}^{L} J_z \sigma_k^z \sigma_{k+1}^{z} + \vec{B} \cdot \vec{\sigma}_k~,
\end{equation}
with $\vec{B}$ a homogeneous magnetic field. Here, since the number of points
in quantum state space is $2^{L_{\mathrm{env}}}$, where $L_{\mathrm{env}}$
is the size of the environment, solving the equations of motion analytically becomes
unfeasible, and we resort to a numerical approach, in which we can directly evaluate
physical quantities of interest, such as the information flux, the sink/source term and
the entropies of the various distribution involved. In our numerical analysis, at $t=0$
the system and the environment begin in a product state, in which the state of the system
is $\ket{1}$. Hence, the geometric quantum state is a dirac delta distribution, concentrated 
on the north pole. However, as time goes by, the overall pure state is not a product anymore
and we end up with a non-trivial geometric quantum state, whose evolution
is dictated by Eq.(\ref{eq:kinetic}) or, at the phenomenological level, by the continuity equation. 
We can map the evolution of the geometric quantum state on the surface of the Bloch sphere, 
as done in Figure \ref{fig:gqs_dynamics} for example, or using $(p,\phi)$ coordinates, on the 
Bloch square. 

\subsection*{Anti-Ferromagnetic example}
In Figure \ref{fig:gqs_dynamics2} 
we show the evolution of the geometric quantum state of a qubit with an environment of 12 other
qubits, all interacting with an Ising Hamiltonian with $J_z=1$, $\vec{B}=(1,0,0.5)$ and 
with an initial uniform product state $\ket{\uparrow \ldots \uparrow}$. Here we can see that
the geometric quantum state, while initially localized, quickly spreads around to reach an 
equilibrium distribution. It is, however, interesting to see that it reaches equilibrium with 
a highly structured fashion. Here we can see that for $t\approx 0$ the distribution of points 
$\Gamma_\alpha(t)$ is organized in three clusters of states which are quite close to each other. 
As time goes by, each of these clusters \emph{tri-furcates} into smaller clusters of less points. 
After this splitting occurs a certain number of time we loose track of the clustering and the distribution reaches
an equilibrium configuration. The origin of this behavior has to be tracked to the structure of the
operators $\hat{V}_{\alpha \beta}$. In particular, be believe this the tri-furcation to be due to the fact 
that there are two nearest neighbors directly interacting with our system. Increasing this number 
would result in a more detailed splitting, increasing the number of sub-clusters. To confirm this, 
we performed the same exact simulation, with a next-to-nearest neighbor Ising model, observing 
the formation of $5$ sub-clusters at each splitting. A phenomenon we can call ``penta-furcation''. 
This suggests the number of new subclusters $n_{\mathrm{sc}}$ to be equal to the number 
$n_{\mathrm{int}}$ of qubits our system interacts directly with: $n_{\mathrm{sc}} = n_{\mathrm{int}}$.
While we do not have a definitive explanation for this phenomenon, it has a distinct, and 
unexpected, fractal flavor which deserves further investigation.

After a transient of $60/70$ time-steps, with $\delta t = 0.005$, we can see that the distribution
settles on an equilibrium configuration that is almost flat. This can be seen by looking
directly at the distribution of information contained in a certain region of the state space.
By discretizing the Bloch square, and therefore $\mathbb{C}P^1$, via its canonical coordinates,
we can integrate the geometric quanutm state on the small cells, track their evolution and
extract a phenomenological evolution equation. 
Given a certain number $N_p N_\phi$ of desired cells  $\mathcal{C}_{ik} = \left[ \frac{i}{N_p}, \frac{i+1}{N_p}\right[ \times \left[ \frac{2\pi k}{N_{\phi}},\frac{2\pi (k+1)}{N_{\phi}}\right[$,
we have $\bigcup_{i=1}^{N_p}\bigcup_{k=1}^{N_{\phi}}\mathcal{C}_{ik} = \PH$, resolution $\delta_p = \frac{1}{N_p}$ and
$\delta_\phi = \frac{2\pi}{N_\phi}$.
Given a geometric quantum state $q_t(Z)$, the probability mass in each cell is $\mu_{ik}(t) = \int_{\mathcal{C}_{ik}}q_t(Z) dV_{FS}$.
By integrating the continuity equation (Eq.(\ref{eq:continuity})) over a cell $\mathcal{C}_{ik}$
we get the following equation:
\begin{equation}
\frac{d\mu_{ik}}{dt} + F_{ik} = \sigma_{ik}~,
\end{equation}
where
\begin{align}
&\sigma_{ik}(t) = \int_{\mathcal{C}_{ik}} \!\!\! dV_{FS} \sigma_t(Z)  \\
&F_{ik} = \int_{\mathcal{C}_{ik}} \!\!\! dV_{FS} \nabla \cdot J_t~.
\end{align}
Using the divergence theorem, we can turn $F_{ik}$ into an integral
over the surface $\partial \CC{i}{k}$ of $\CC{i}{k}$:
\begin{equation}
F_{ik} = \int_{\partial \mathcal{C}_{ik}} \!\!\!  J_t \cdot n dS_{FS} = \Phi_{ik}(J_t)~,
\end{equation}
where $n=(n_p,n_\phi)$ is the vector orthogonal to the boundary of $\CC{i}{k}$
and $dS_{FS}$ is the surface element. This quantity, $F_{ik}=\Phi_{ik}(J)$ is 
the flow of probability mass across the boundary of each cell $\CC{i}{k}$
due to the underlying movement of the points $\Gamma_\alpha(t)$. On the other
hand, $\sigma_{ik}(t)$ quantifies the probability mass that is produced or sinked
in $\CC{j}{k}$. Beyond the initial transient, as the distribution settles on a more
regular behavior, we expect a more regular evolution equation to emerge. In
particular, we expect two emergent behavior. First, we assume the sink/sources
term to be zero after a short transient $\sigma_{ik} \approx 0$. This assumption is
motivated by two physical reasons. Since the $x_\alpha(t)$ are the probabilities of 
finding the environment in one of the elements of its eigenbasis $\ket{b_\alpha}$, 
for a large environment (the number of points is exponential in the size of the system) 
it is reasonable to expect that after a short transient these probabilities will settle on 
``equilibrium'' values (as the ones given by max entropy) and, possibly, oscillate 
around them. Second, we are interested in looking at the system on longer time-scales, 
where these oscillations can be averaged out. Together, these arguments supports the
assumptions for $\sigma_{ik}\approx 0$. This implies that the evolution is guided only 
by the flux term, leading to a local conservation of probability mass. Thus, we can 
now build a microscopic, particle-like, model for $\Phi_{ik}(J_t)$.
Since probability mass is locally conserved, the only way $\CC{i}{k}$ can loose or 
gain probability mass is by exchanging particles with its immediate neighbors: $\CC{i\pm1,}{k}$ and $\CC{i,}{k\pm1}$. 
Thus, as first approximation, we assume there is a fixed rate of gain and loss of particles,
in each direction $(p,\phi)$: $\gamma_p$ and $\gamma_{\phi}$. This leads to 
the following phenomenological model for the evolution of $\mu_{ik}(t)$:
\begin{align}
\frac{d\mu_{ik}}{dt} &= \gamma_p \left( \mu_{i+1,k} + \mu_{i-1,k}\right) +\gamma_{\phi} \left( \mu_{i,k+1} + \mu_{i,k-1}\right) \nonumber \\
&- \gamma_{\mathrm{loss}}\mu_{ik}~.
\end{align}
The first two terms represent the gain due to a certain number of particles entering 
from each of the four sides of a cell $\mathcal{C}_{ik}$, either from $p$ or $\phi$ direction. 
The last term represents the loss due to a certain number of particles leaving $\CC{i}{k}$ 
for one of its four adjacent neighbors. By assuming that there are no special cells, unless otherwise 
specified, this leads to $\gamma_{\mathrm{loss}} = 2(\gamma_p+\gamma_{\phi})$.
This model leads to a simple diffusive behavior that, while not entirely accurate, 
clearly captures the essence of the evolution beyond the transient. While finite-size
fluctuations in time are to be expected, as we increase the size of the Ising environment
we expect them to be less relevant, and our model more accurate. A full analysis of 
the phenomenology of this model will be carried elsewhere. Here we are interested 
in proving that this theoretical framework can lead to tractable phenomenological 
models. Indeed, in the limit of an infinitely fine-grained discretization this model leads 
to a Fick's law of information diffusion $\PH$, as $J^{\mathrm{diff}}_t =  (-\gamma_p \frac{\partial q_t}{\partial p},-\gamma_\phi \frac{\partial q_t}{\partial \phi} ) = - \gamma \cdot \nabla q_t$
and therefore to a diffusion equation in $\PH$:
\begin{equation}
\frac{\partial q_t}{\partial t} =  \gamma \cdot \nabla^2 q_t~,
\end{equation}
where $\gamma = (\gamma_p,\gamma_\phi)$ are direction-dependent diffusion coefficients.
\begin{figure}[t!]
\centering
\includegraphics[width=.3\textwidth]{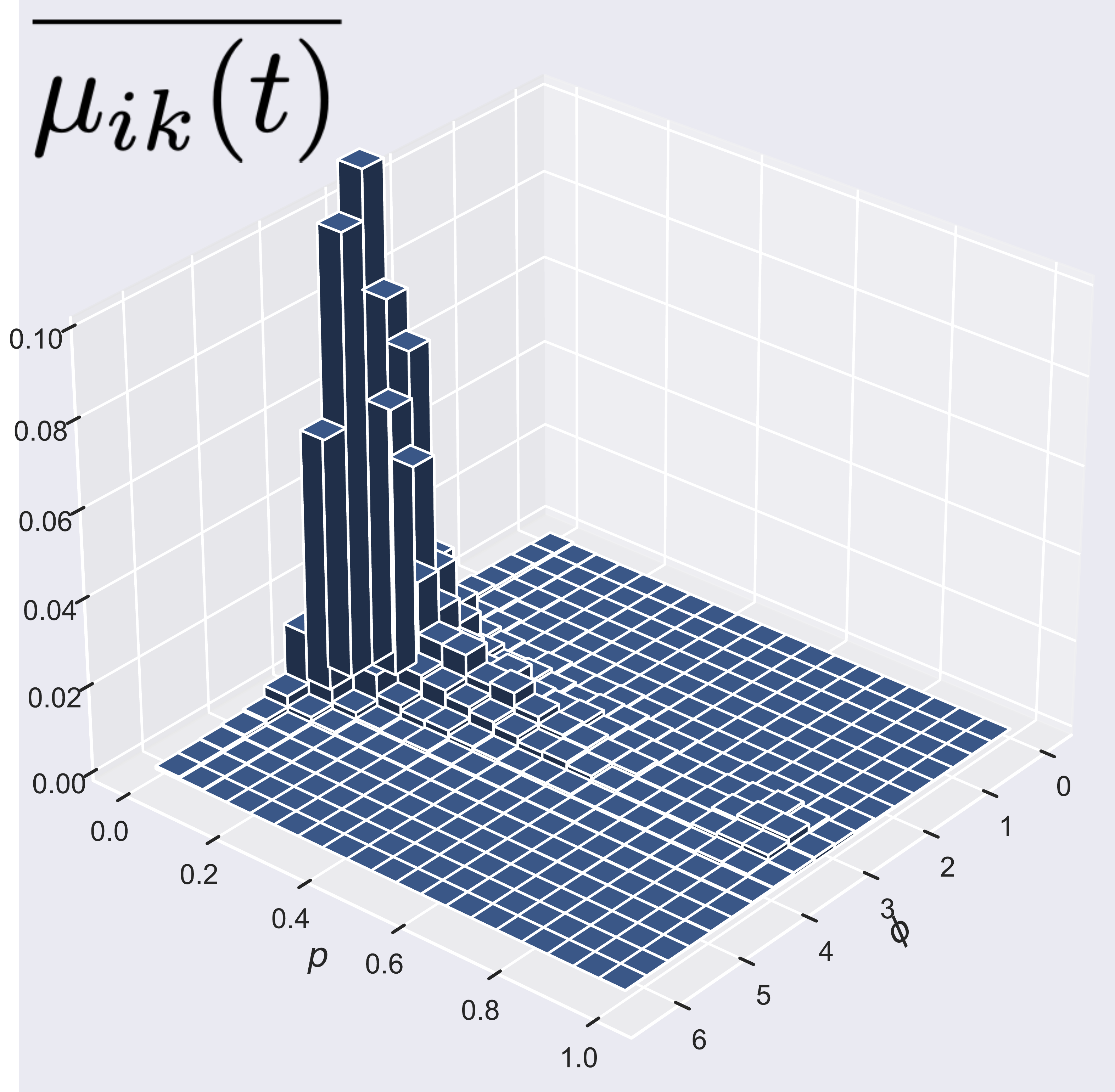}
\caption{Time-average of the coarse-grained geometric quantum 
state, $\mu_{ik}(t)$, which conveys the information about how much
information about the system is contained in a small cell $\CC{i}{k}$ of state space.
In the case of a ferromagnetic Ising environment the non-equilibrium behavior
of the quantum system maintains forces the information to be localized in a certain
region of the state space, loosely localized around $Z_{center}=(p=0.1,\phi = \pi)$.
Here this is shown by plotting the time-average of $\mu_{ik}(t)$, after the first
$100$ time-steps, to avoid including transient behavior. A localized distribution
is evident.
	}
\label{fig:mu_average}
\end{figure}
\subsection*{Ferromagnetic example}
In figure \ref{fig:gqs_dynamics3} we show the evolution of the geometric
quantum state for a qubit in a ferromagnetic Ising environment. We do not
show the short-time transient, as it is qualitatively similar to the antiferromagnetic case,
still exhibiting the tri-furcation dynamical pattern. Rather, we focus on two aspects
of this non-equilibrium evolution, which were absent from the antiferromagnetic example: \emph{sustained 
fluctuations}. We have highlighted them with the black arrow superimposed to the top left
plot, and are evident in the video, which can be found at
\url{https://github.com/fabioanza/GeomQuantMech/blob/main/video_g1_h05_L12_ferro_long.mp4}.
At its core, this is an emergent phenomenon in which the probability mass, instead of
diffusing across the whole state space as in the antiferromagnetic example, appears to be 
rotating around a point loosely localized as $Z_{\mathrm{center}} \approx (p=0.1,\phi=\pi)$. 
As a result of this, over time the distribution $\mu_{ik}(t)$ appears to remain localized in 
a small region around $Z_{\mathrm{center}}$. This can be seen by plotting the time-averaged
information distribution $\overline{\mu_{ik}(t)} = \frac{1}{400}\sum_{n=101}^{500} \mu_{ik}(t_n)$, 
in which the time-average has been performed by excluding the first $100$ time-steps, to 
exclude the transient behavior. The result is shown in the two bottom rows of Figure \ref{fig:summary}.  We now 
use the approach developed in the previous section to understand how this occurs. 
Indeed, by plotting the coarse-grained flux of information $J_{ik}(t) = \int_{\CC{i}{k}}J_t(Z) dV_{FS}$
and the coarse-grained information production term, in Figure \ref{fig:summary} we see that 
the localization of the information around $Z_{center}$, and its stability over time, arises 
from the competition between a sink term, drawing up information and delocalizing it across
the state space, and a rotating vector field $J_{ik}(t)$, moving the points away from the sink 
but maintaining them close to $Z_{center}$, with an overall rotation effect. Over time, these two
effects balance each other out maintaining the information localized around $Z_{center}$, 
as it can be seen from the instantaneous distribution $\mu_{ik}(t)$, plotted in the top
row of Figure \ref{fig:summary}.

While a fully analytical description of this phenomenon requires an in-depth investigation of the 
microscopic details, the analysis given in term of information flux and information sink and source
term provides a clear connection between the microscopic, kinetic, evolution and the macroscopic
behavior provided by expectation values of observables like $\MV{\sigma_z}$.

\section{Conclusions}
\label{sec:FINAL}

We have developed a consistent framework to study how quantum systems move
information around their state space. Our main results are the continuity equation (Eq.(\ref{eq:continuity}))
and the underlying kinetic equations (Eq.(\ref{eq:kinetic})). Together, they regulate how information
can be transported around the quantum state space as a result of the interactions between the
system and its environment. The kinetic equation represents the microscopic, detailed, model 
while the goal of the continuity equation formulation is to extract emergent, simplified, phenomenological descriptions. 
As an example of this logic, in Section \ref{sec:EXAMPLES1} and \ref{sec:EXAMPLES2} 
we have looked at the transport of information in the state space of a qubit. In both cases we have 
seen the emergence of consisten behavior that can be described with sufficiently simple information 
fluxes. In the case of an isolated qubit, we were able to analytically extract the information flux. The 
case of an antiferromagnetic Ising environment with both longitudinal and transverse field led us to 
see how a simple diffusion model can reproduce the macroscopic feature of the kinetic evolution.

While here we focused only on the theoretical framework, we believe we have shown that the 
study of concrete kinetic models of information transport is phenomenologically quite rich, and
deserves to be investigated in depth. The underlying reason behind our efforts in this kind of geometric-oriented 
research direction is that we believe these new tools will allow us to (1) gather a more accurate understanding
and (2) build improved models for the nonequilibrium behavior of quantum systems. In particular, we
believe these tools will reveal to be appropriate to get a more accurate picture of the interplay between 
the non-equilibrium physical properties of open quantum systems and the information-theoretic features
of their dynamics, which emerge from their complex stochastic nature.

Future work in this direction will be focused on extracting more general, approximated, analytical models and 
developing more powerful numerical techniques, to investigate the behavior of quantum systems with larger 
environments. An interesting concrete issue is to understand how quantum information moves around the 
quantum state space when our system is part of a quantum computer performing quantum computation.
Indeed, we believe this framework can help better understand how to improve and optimize quantum information
processing for future quantum technologies.

We conclude with a general comment, clarifying our main goal behind the paper, and with a series of future directions
opened by our work. While our main goal here was to develop the theoretical framework and build the intuition to 
support it, we believe the general analyses presented in Sections \ref{sec:EXAMPLES1} and \ref{sec:EXAMPLES2} 
revealed four features of interest, which deserve attention on their own. First, the dynamical emergence of the 
fractal pattern in the transient behavior of the antiferromagnetic model was unexpected and one wonders
if similar dynamical patterns are typical of quantum models or not. This is clearly related to the Second point: 
the more general question of describing stationary states, with non-trivial non-equilibrium behavior.
A natural route starts with imposing $\frac{\partial q_t}{\partial t} = 0$ in the continuity equation, which in turn 
leads to a balance equation between information flux and information sinks and sources. While one
can expect that $J_t=\sigma_t=0$ characterizes thermal equilibrium, this will not be true in general, and the
continuity equation can be used to characterize stationary states. Third, in the Discussion part of Section \ref{sec:EXAMPLES1} (see 
Eq. (\ref{eq:nochaos})), we have seen that a unitary dynamics is a very rigid evolution in which there is a unique
velocity field (a divergence-free one) and all points in the support of any geometric quantum state have to 
move in the same way. It is well known from the theory of dynamical systems that this is not a feature of all
Hamiltonian evolutions, which are able to support chaotic phenomenology via stretching and folding. Therefore, 
a quantum state space should be able to support more general Hamiltonian evolutions, exhibiting richer phenomenology
as chaos, of quantum nature. Thus, we believe the geometric formalism and the framework developed here to 
be a concrete route to improve our understanding of quantum chaos. Fourth, in Section \ref{subsec:IQS} we have 
proven a novel form of Liouville's theorem for quantum systems. This poses a conundrum as it is in stark contrast with 
all other phase space formulation of quantum mechanics, such as Wigner's \cite{Wig32} or variations of it such as 
the one based on the Husimi function \cite{Hus40}, both of which are known to violate Lioville's theorem \cite{Oliva18}. 
The validity of Liouville's theorem is not just a mathematical fact, but a truly important property of the dynamical 
evolution, which can help us characterize useful properties of the dynamics, such as the emergence of chaos. 
Here, again, the geometric formalism and the theoretical framework developed here is bringing to light interesting
new insights into the nonequilibrium behavior of quantum systems.

\section*{Acknowledgments}
\label{sec:acknowledgments}

We thank James Crutchfield, David Gier, Ariadna Venegas-Li and Dhruva Karkada
for discussions on the geometric formalism of quantum mechanics and the Telluride
Science Research Center for its hospitality during visits.  This material is
based upon work supported by, or in part by, a Templeton World Charity
Foundation Power of Information Fellowship.

\newpage
\bibliography{library}

\end{document}